\documentclass[prl,twocolumn,superscriptaddress]{revtex4}
\usepackage{amsmath,amssymb,mathrsfs}
\usepackage[dvips]{graphicx}
\usepackage{hyperref}
\usepackage{paralist}

\def\be{\begin{equation}}
\def\ee{\end{equation}}
\def\bea{\begin{eqnarray}}
\def\eea{\end{eqnarray}}

\def\tbf{\textbf}

\begin{document}

\title{Bose-Einstein supersolid phase for a novel type of
momentum dependent interaction}

\author{Xiaopeng Li}
\affiliation{Department of Physics and Astronomy, University of Pittsburgh, Pittsburgh, Pennsylvania 15260, USA}

\author{W. Vincent Liu}
\email[e-mail:]{w.vincent.liu@gmail.com}
\affiliation{Department of Physics and Astronomy, University of
Pittsburgh, Pittsburgh, Pennsylvania 15260, USA}
\affiliation{Center for Cold Atom Physics, Chinese Academy of
Sciences, Wuhan 430071, China}

\author{Chungwei Lin}
\affiliation{Department of Physics and Astronomy, University of Pittsburgh, Pittsburgh, Pennsylvania 15260, USA}

\date{\today}

\begin{abstract}
A novel class of non-local interactions between bosons is found to
favor a crystalline Bose-Einstein condensation ground state. By using
both low energy effective field theory and variational wavefunction
method, we compare this state not only with the homogeneous
superfluid, as has been done previously, but also with the normal
(non-superfluid) crystalline phase and obtain the phase diagram.  The
key characters are: the interaction potential displays a negative
minimum at finite momentum which determines the wavevector of this
supersolid phase; and the wavelength corresponding to the momentum
minimum needs to be greater than the mean inter-boson distance.

%{\bf 
%A system of bosons with non-local interaction is studied. From the effective field %theory 
%of low energy excitations and energy comparisons of several variational states, a %novel 
%class of interaction potentials is found to favor crystalline Bose-Einstein %condensation 
%ground state against both the standard homogeneous superfluid and non-superfluid 
%crystalline states. The key character is that the potentials display a negative %minimum at a finite
%momentum which determines the wave vector of the crystalline Bose-Einstein %condensation.
%}  
\end{abstract}

\maketitle 

Since Penrose and Onsager's first discussion~\cite{PO} on
existence of ``supersolid'', namely a phase with co-existence of
superfluid and crystalline order, both experimental~\cite{expssbundle}
and theoretical~\cite{ssbundle} attempts have been made for decades in
the search of this novel phase.  Recently reported observation of
``supersolid'' phase in He-4 systems~\cite{EKss} revitalized this
fundamental interest. Nevertheless, some subsequent experimental
evidences as well as various proposed microscopic
mechanisms~\cite{Prokofsupersolid} remain controversial. 

Progress on the physics of cold atoms and molecules opens a new
possibility to study the ``supersolid'' phase thanks to clean and
controlled experimental systems. One of the most fascinating facts is
the unprecedented tunability of the interaction potentials due to
internal degrees of freedom of atoms and molecules~\cite{MicheliMol,Pupillo_Rydberg}, 
which allows one to address a theoretical question,
namely what interaction potentials can support the supersolid phase in
continuous space. Recent experimental progress on dipolar quantum
gases allows to explore new physics of quantum many body systems with
non-local interactions~\cite{KRbbundle}.  It is well established that
non-local interaction potentials stabilize the supersolid phase on
lattice~\cite{dipolarreview}.  The possibility of finding a
Bose-Einstein supersolid phase~\cite{PomeauBES} was also put forward
for several continuum model systems such as dipolar quantum
gases~\cite{dipBES}, atom-molecule mixture gases~\cite{AMSFRadzi} and
Rydberg atom gases~\cite{HenkelBES,Pupillo_Rydberg}.  
Recently,
Henkel {\it et al}~\cite{HenkelBES} found that the Fourier transform
of an isotropically repulsive van der Waals interaction potential with
a ``softened'' core has a partial attraction in momentum space, which
gives rise to a transition from a homogeneous Bose-Einstein condensate
to a supersolid phase due to roton instability.
%% is self-consistently obtained within Gross-Pitaevskii theory. 
 However,
whether the supersolid phase they found 
is stable against fluctuations and
how  it should compare with the non-superfluid (normal) crystal phase
has  not been studied.
Recent work on dipolar gases~\cite{KaushikDip} showed that the dipolar
dominating interaction does not support a supersolid phase in the
phase diagram between the uniform superfluid and the normal crystal
phase, where this phase had been speculated to exist.
%What interaction potentials can 
%support and stabilize the supersolid phase and what parameter regime 
%one should look for is still an open question.  

\begin{figure}[htp]
\includegraphics[angle=0,width=0.8\linewidth]{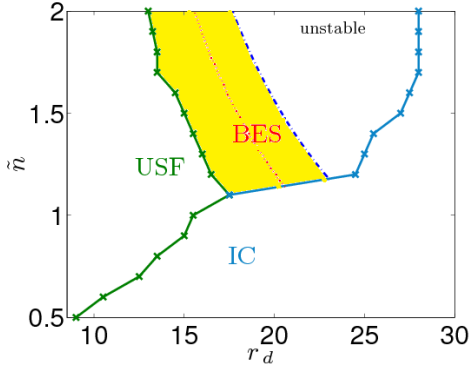}
\caption{The phase diagram of bosons with step-like
  interaction. Uniform superfluid (USF), insulating crystal (IC), and
  Bose-Einstein supersolid (BES) phases are separated by `solid lines'
  with ``$\times$'' showing the data points from variational
  calculation.  Analysis of the collective excitation spectrum shows
  the instability of USF at the `red dotted line' and that of BES at
  the `dark blue dash dotted line'. At low density, USF phase and IC phase 
exist; while at high density the new stable BES phase is found in the 
 `yellow shaded' regime. When $\tilde{n} \gtrsim 1$, the IC state is not stable 
(see text). 
}
\label{PhaseD}
\end{figure}

In this Letter, we show that interaction potentials, which display a
minimum of negative value at a finite momentum, lead to a
modulating superfluid order, namely a Bose-Einstein supersolid (BES)
phase. {We perform effective field theory analysis and variational
calculation to determine not only the phase boundary between the
uniform superfluid (USF) phase and BES, which has been previously
analyzed by roton instability for dipolar~\cite{PomeauBES,dipBES} or
van der Waals interaction~\cite{HenkelBES}, but also the phase
boundary between BES and the normal (non-superfluid) insulating
crystal (IC) phase. We shall begin with a heuristic argument to show
how a stripe BES phase should arise from the competition between kinetic
and interaction energy
in the regime of roton instability.  
Next, we shall study as a concrete example the ``softened'' dipolar 
%% step-like
interaction recently proposed for Rydberg atomic
gases~\cite{Pupillo_Rydberg}.
We will establish the ground state in the sense
of variational principle and find a first order phase transition from
the uniform superfluid phase to the triangular crystalline BES phase.
Finally, we shall compare the energies of BES and IC phases of the same
lattice configurations, and find a regime in which the
triangular-lattice BES is stable and has lower energy than both USF
and (normal)
IC. The result is summarized in Fig.~\ref{PhaseD}.}

\paragraph{Hamiltonian and heuristic.} 
To explore the physics of the BES phase, we start with the
continuum Hamiltonian of two dimensional interacting bosons 
%a modified purely  repulsive dipolar interaction,
\bea
&&H=\int d^2 \vec{r} \hat{\psi}^\dag (\vec{r})\left[-\frac{\hbar^2}{2m}\nabla^2-\mu\right] \hat{\psi}(\vec{r})   \nonumber       \\
&+& \frac{1}{2}\int d^2\vec{r}_1 d^2 \vec{r}_2 \hat{\psi}^\dag(\vec{r}_1) \hat{\psi}^\dag(\vec{r}_2)
		V(\vec{r}_1-\vec{r}_2)
		\hat{\psi}(\vec{r}_2) \hat{\psi}(\vec{r}_1), \, ~~
\eea
where the first term of $H$ corresponds to the kinetic energy, and the second  the 
two-body interaction energy.

It is commonly accepted that the ground state for such a continuous bosonic system should be USF at
the kinetic energy dominating regime. The USF phase is described by a coherent state 
$  |\text{USF} \rangle = \exp(\int d^2 x\sqrt{n} e^{i \phi_0} \hat{\psi}^{\dag}(x)) |\Omega \rangle $ 
where $n$ is the mean particle density, $\phi_0$ a constant phase, and $|\Omega \rangle$ the vacuum state 
with no particle.
The energy of this state is given by $E_{\text{USF}} = \frac{N}{2}nU(\tbf{k}=\textbf{0})$, where 
$N$ is the mean particle number and $U(\tbf{k})$ is Fourier transform of 
the interaction potential.
We first analyze the instability of the USF phase. This can be 
performed using an effective field theory approach~\cite{Popov,Xiaogang}. The real
 time action of this bosonic system is  
$S[\bar{\psi},\psi]= \int d^{2} x \,dt \left\{i \hbar \bar{\psi} \partial_t \psi -\mathscr{H}[\bar{\psi},\psi] \right\}  $.
Fluctuations on top of the uniform superfluid state are considered by 
writing the boson field $ \psi(x,t)= [\rho_0 + \delta \rho]^{1/2} e^{i\phi}$, 
assuming $\delta \rho$ and $|\nabla \phi|$ are small.  
%The effective theory for small fluctuations is described by the following action
%\bea
%S_{\mathrm{eff}} [\delta \rho, \phi]& =&\int dt \int d^2 x \mathscr{L} (x,t)
% \nonumber \\ 
%\mathscr{L} &=& -\delta \rho \hbar  \partial_t \phi
%                  -\frac{1}{8\rho_0}(\hbar \vec{\nabla} \delta \rho)^2
%                            -\frac{1}{2} \rho_0 (\hbar \vec{\nabla}
% \phi)^2 \nonumber \\ 
%                            &-&  \frac{1}{2} \int d^2 x' V(x-x')
% \delta \rho(x,t) \delta \rho(x' , t)\, , \, ~~
%\eea
%where $\rho_0$ denotes the condensate density which is a constant in space.
%Expanding the action to second order, we can get a quadratic effective action for 
%fluctuations. 
The quasiparticle spectrum is readily derived after integrating out
the $\delta \rho$ field: $\epsilon(\textbf{k})=\sqrt{\frac{\hbar^2
\textbf{k}^2}{2m}(\frac{\hbar^2 \textbf{k}^2}{2m}+2 n
U(\textbf{k}))}$. For a potential that has a negative minimum at a
finite momentum, this spectrum at that momentum drops,
eventually hits zero and becomes imaginary when increasing the density
$n$. That suggests that the assumed USF
(coherent) state is unstable towards possible crystalline order.

To show the BES phase arises, we first give a heuristic argument by considering a 
simple stripe BES state
$|\text{BES}\rangle = \exp \left (\sqrt{N} (\frac{\sqrt{2}}{2} b_{\textbf{Q}/2} ^{\dag}
+\frac{\sqrt{2}}{2} b_{-\textbf{Q}/2} ^{\dag}) \right) |\Omega \rangle $, where
$\textbf{Q} = [Q,0]$, and $\textbf{Q}$ the minimum point of $U(\textbf{k})$. The 
energy of this state is given by 
$E_{\text{sBES}} = N \left(\frac{\hbar^2 Q^2}{8 m }+\frac{1}{4} n U(\textbf{Q})\right)+E_{\text{USF}}$. 
When the term
$ \frac{\hbar^2 Q^2}{8 m }+\frac{1}{4} n U(\textbf{Q})$ is negative, namely the interaction energy 
dominates over the kinetic energy, the stripe BES state has 
lower energy than the USF state. (We also go beyond the mean field state and 
compare with the two component fragmented state 
$ |f \rangle = \sum_{l= -N/2} ^{l=N/2} \alpha_l\frac{(b_{\textbf{\tbf{Q}}/2} ^\dag) ^{\frac{N}{2} +l}
(b_{-\textbf{\tbf{Q}}/2} ^\dag) ^{\frac{N}{2} -l} } {\sqrt{(\frac{N}{2}+l)! }
\sqrt{(\frac{N}{2}-l)! } } |\Omega \rangle $, where $\{\alpha_l\}$ are variational 
parameters~\cite{MuellerFragmentation}, and the coherent stripe BES state is found to have the lowest 
energy.) We thus conclude 
the BES state arises from the competition of kinetic energy and interaction energy.

To be concrete, we further apply the two-particle interaction of a step-like form
$V(\vec{r}) = \frac{D}{r_0 ^3} $ if $r<r_0$; $V(r) = \frac{D} {r^3}$ otherwise.
The form of this potential is an approximation to the interaction between 
polarized Rydberg atoms  
proposed in Ref.~\cite{Pupillo_Rydberg}.
Two dimensionless 
parameters of this system are  $\tilde{n}\equiv n\times r_0^2$ and $r_d\equiv\frac {2mDn^{1/2} }
{\hbar^2}$. $\tilde{n}$ characterizes the relation between $r_0$ and the inter-particle 
distance, and $r_d$ characterizes the strength of interaction.
%We aim at finding the phase diagram of this system parameterized by $\tilde{n}$ and $r_d$ at zero %temperature.
A phase transition from
USF to IC has been found when varying $r_d$ at the regime 
of $\tilde{n}\approx 0.9$~\cite{KaushikDip}. 
The IC (single particle per site)
phase is described in a second quantization form by $
|\Psi_{\text{IC}}\rangle=\prod_{\vec{R}_i} {c_{\vec{R}_i}^\dag}|0\rangle $,
where $\vec{R}_i$ is the  direct lattice vector at site $i$,
and the single particle wavefunction corresponding to
${c}^\dag_{\vec{R}_i}$ is the Wannier function $\phi_{\vec{R}_i}
(\vec{r})$.

The Fourier transform of this step-like interaction is shown in 
FIG.~2(a). %~\ref{Spectrum}
It is straightforward to obtain the excitation spectrum, which is shown 
in FIG.~2(b). %\ref{Spectrum}(b). 
It can be seen that the spectrum displays instability. The origin of this
effect is that the Fourier transform of the interaction, $U(\tbf{k})$, has a
negative minimum at a finite momentum. 
Now the question is to find the stable variational minimum in the coherent state space. 
With $|G \rangle = \exp (\int d^2 \textbf{x} \phi(\textbf{x}) \hat{\psi} ^{\dag} 
(\textbf{x})) |\Omega \rangle$ (so that 
$\hat{\psi}(\textbf{x}) |G \rangle = \phi(\textbf{x}) |G \rangle $), 
the energy of this state is readily given by: 
%Now the question is to find the condensate wavefunction which minimizes  
%the GP energy functional.
%The GP energy functional for condensate
%wavefunction $\phi(\vec{r})$ is: 
\bea
E[\phi,\phi^*]&=&\int d^2 \vec{r}\frac{\hbar^2}{2m}|\vec{\nabla}\phi|^2 \nonumber \\
&+&\frac{1}{2}\int d^2\vec{r}_1
d^2\vec{r}_2V(\vec{r}_1-\vec{r}_2)|\phi(\vec{r}_1)|^2|\phi(\vec{r}_2)|^2
\,, \, ~~
\eea
where $V(\vec{r})$ is the interaction potential.

\begin{figure}
\includegraphics[width=0.45\linewidth]{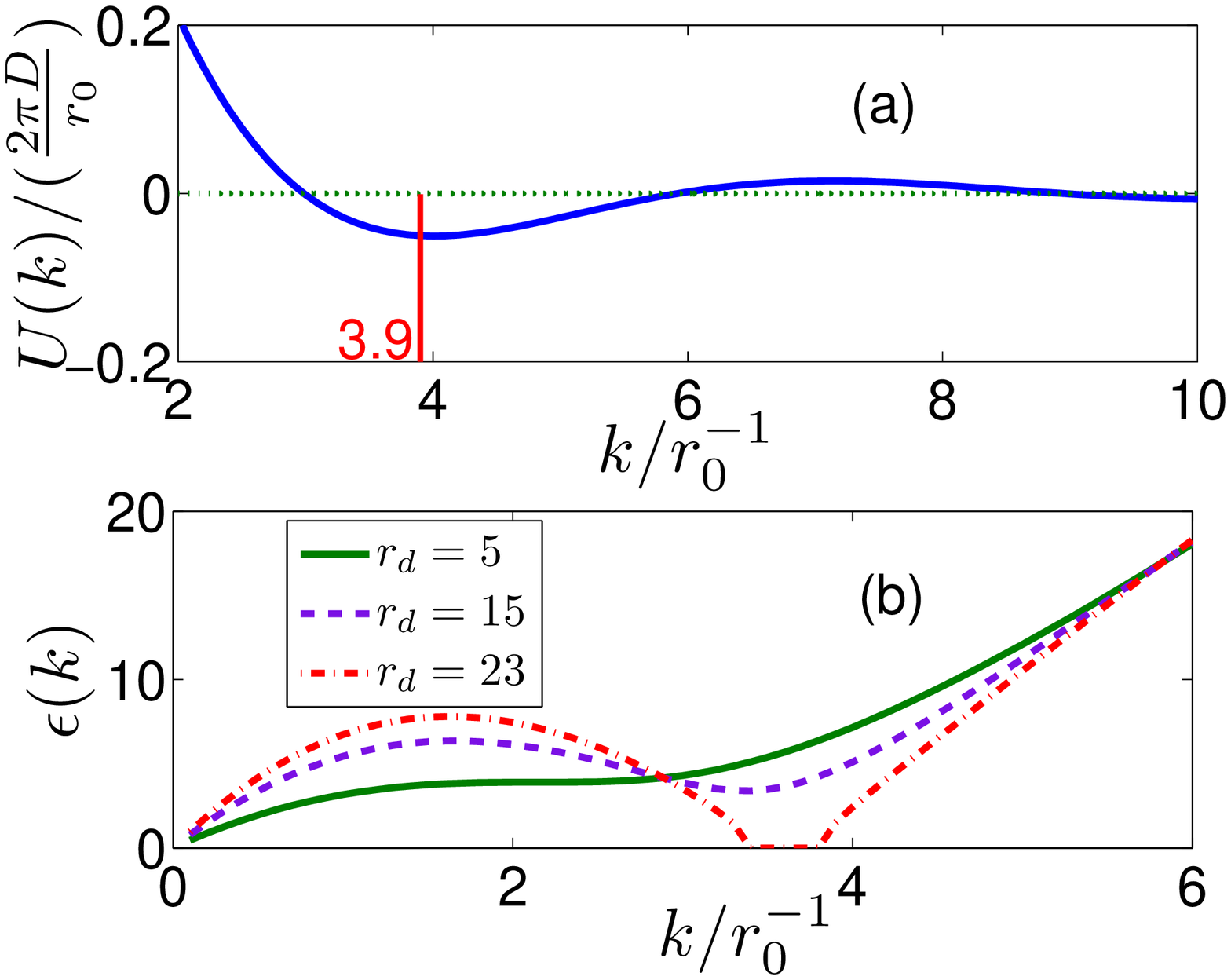}
\includegraphics[width=0.52\linewidth]{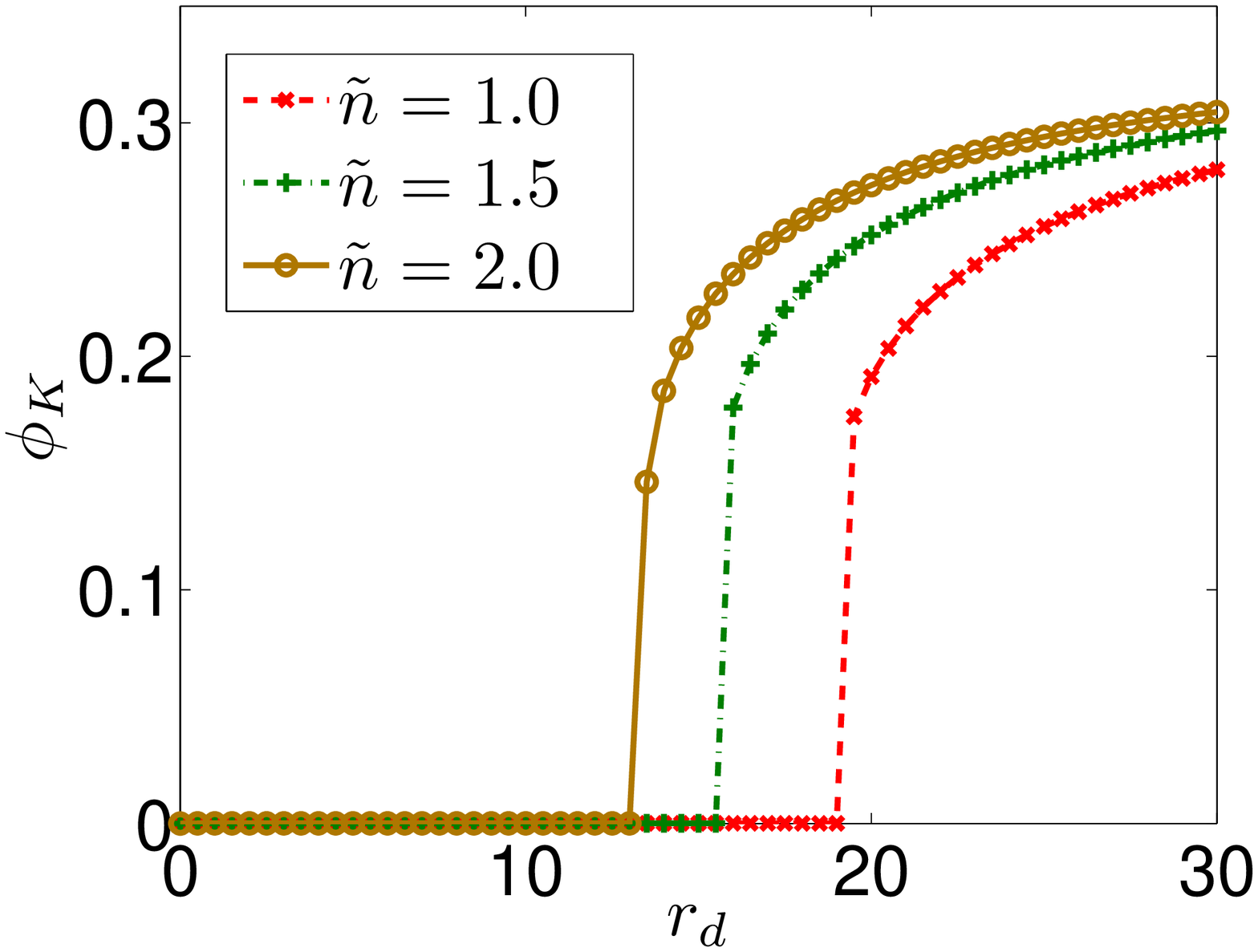}
\caption{LEFT figure~(a): Fourier transform of the step-like two-body interaction.
LEFT figure~(b):\label{Spectrum} Shows the Bogoliubov quasiparticle spectrum for
a USF state. The unit of $\epsilon_k$ is ${\hbar^2}/(mr_0^2)$. The
plot shows the real part of the spectrum with $\tilde{n} =1$. The solid line
corresponds $r_d=5$, the dashed line to $r_d=15$, and the dashed
dotted line $r_d=23$.  RIGHT figure: \label{PT} Shows the phase transition from
the USF to the triangular crystalline BES phase.
 $|\phi_{\textbf{K}} |^2 \equiv \frac{1}{N}\langle b ^{\dag}_{\textbf{K}}  
b_{\textbf{K}} \rangle$ is the 
occupation fraction of the lowest finite momentum.} 
\end{figure}

\paragraph{ Variational analysis.}
We first check whether the system favors an extended or localized state. 
This purpose is fulfilled by applying the Gaussian ansatz which
means $\phi(\vec{r})=\frac{\sqrt{N}}{\sqrt{\pi} \sigma} e^{-\frac{|\vec{r}|^2}{2\sigma^2}}$. The total energy of
this system is given by $E_t=E_k+E_{dip}$, where the kinetic energy
$E_k= N\frac{\hbar^2}{2m\sigma^2}$ and the interaction  energy
$E_{dip}= N^2 \frac{1}{2 \pi^2} \frac{D}{r_0 ^3}
g(\frac{r_0}{\sigma})$. Here, $ g(x) $ is approximately $ \pi^2 (1-e^{-2x^2}) $. 
 The energy per particle
 is  $\frac{\hbar^2}{2m\sigma^2}+\frac{ND}{2\pi^2 r_0^3}g(\frac{r_0}{\sigma})$. 
In thermodynamic limit $N \rightarrow \infty$, interaction dominates and
$E(\sigma)=\frac{ND}{2\pi^2 r_0^3}g(\frac{r_0}{\sigma})$.
We found that as long as $\int d^2 \vec{r}V(\vec{r})>0$,
 $\sigma \rightarrow \infty$ minimizes the energy, implying
 the system favors an extended state in space.
Since for $r_d>0$ $\int d^2 \vec{r}V(\vec{r})>0$,
we  conclude that the system favors an extended state 
when $r_d$ is positive. 

Up to this point, what we know about this system is that it favors an
extended state which is  not 
  necessarily uniform superfluidity.
%{\bf a uniform superfluid.}
% superfluid state.  
Having argued heuristically above that a momentum dependent interaction may
 favor a finite momentum BEC, it is natural to compare the energy of a new, 
non-uniform coherent state which has discrete lattice symmetries. 
Thus, we can
write the condensate wavefunction in such a form
$\phi(\vec{r})=\langle \psi (\vec{r}) \rangle = \sqrt {n} \sum
_{\textbf{K}}{\phi_{\textbf{K}} e^{i\textbf{K}\cdot \vec{r}}}$ with $\textbf{K}
= p\textbf{G}_1 +q \textbf{G}_2$, where $\textbf{G}_1$ and $\textbf{G}_2$ are 
two primitive vectors spanning the
two dimensional reciprocal lattice.  The corresponding ground state
is $ \left| G \right \rangle =\exp\left( \sum_{\textbf{K}}{\sqrt{N}
\phi_{\textbf{K}} b^{\dag}_{\textbf{K}}} \right) \left |\Omega \right
\rangle $. The order parameter that characterizes the phase transition
from uniform superfluidity to BE supersolidity is an occupation
fraction at some finite momentum $K$, $|\phi_{\textbf{K}}|^2 = 
\frac{1}{N} \langle  b^{\dag}_{\textbf{K}} b_{\textbf{K}} \rangle $.
	
In this assumed ground state subspace, the energy per particle is given by: 
%In this assumed ground state subspace, the expectation value of Hamiltonian divided by $N$ gives the reduced
%GP energy functional which is
\bea
E[\phi_{\textbf{K}}^{*},\phi_{\textbf{K}}]&=&\sum_{\textbf{K}}{\frac{\hbar^2 K^2}{2m}
 \phi_{\textbf{K}}^{*}\phi_{\textbf{K}}} \nonumber \\
&+&\frac{1}{2} n
\sum_{\textbf{K}_1,\textbf{K}_2,\textbf{q}}
{U(\textbf{q}) \phi_{\textbf{K}_1+\textbf{q}}^{*} \phi_{\textbf{K}_2-\textbf{q}}^{*} \phi_{\textbf{K}_2}
\phi_{\textbf{K}_1}}\,. \, ~~
\eea
Now, the problem reduces to minimizing this energy functional with
such a constraint
$\sum_{\textbf{K}}{\left|\phi_{\textbf{K}}\right|^2}=1$, which is equivalent
to the enforcement of conservation of the total particle number. 
By 
${\delta\left(E-\mu \sum_{\textbf{K}}{\phi_{\textbf{K}}^{*}\phi_{\textbf{K}}}\right)}/({\delta
  \phi_{\textbf{K}}^{*}})=0$, we obtain
\be
\mu\phi_{\textbf{K}} =\frac{\hbar^2 K^2}{2m} \phi_{\textbf{K}}+n\sum_{\textbf{K}^{'},\textbf{q}}
{V(\textbf{q})
  \phi_{\textbf{K}^{'}+\textbf{q}}^{*}\phi_{\textbf{K}^{'}}\phi_{\textbf{K}+\textbf{q}}}\, , \,~~
\ee
where $\mu$ is the chemical
potential. 

We compute the energies for three different configurations --- stripe,
square and triangle lattices --- and found that the triangular lattice
is the most  energetically favored. The optimal lattice constant $a_{\text{BES}}$ is
found to be slightly larger than $2 \pi/Q_{\text{min}}$ where $U(Q_{\text{min}})$
corresponds to the negative minimum of the potential. For the particular step-like 
interaction, $Q_{\text{min}}$ is related to $r_0$ by 
$Q_{\text{min}} \approx \frac{3.9}{r_0}\sim \frac{\pi} {r_0}$.
The transition
between the uniform superfluid and the supersolid lattice is of first
order as shown in FIG.~\ref{PT}.

%We first tried the Bogoliubov approach~\cite{LandauStat} 
%of deriving the excitation spectrum. Due to {\bf the strong} dependence on chemical potential of 
%this approach, it is inconvenient to establish the instability boundary of the BES phase.
%We eventually adopted the effective field theory approach to derive the quasi particle spectrum 
%of this BES phase~\cite{Xiaogang, Popov}. 

%
%To study the stability of the BES phase, we further explore
% fluctuations on top of the ground state.  We first tried the
%Bogoliubov approach \cite{LandauStat} of deriving the excitation 
%spectrum. The difficulty of this approach is how to put in
%the constraint due to conservation of particles. The simplest way of
%dealing with this constraint would be to introduce the chemical
%potential. But we found the result of this approach eventually becomes
%sensitive to the chemical potential, and the physics is not quite
%clear either.  We eventually adopted the effective field
%theory approach to solve this problem~\cite{Xiaogang}.
%}
\paragraph{Dynamic stability of the Bose-Einstein supersolid phase.}
To study the stability of the BES phase, we further explore the fluctuations 
on top of the ground state. 
In the presence of BES phase, we can get the effective field
theory for the density and phase fluctuations~\cite{note1},
 $\delta \rho (\tbf{x},t)$ and
$\varphi(\tbf{x},t)$, respectively, writing
$\psi(\tbf{x},t)=[\rho_0 (\tbf{x}) +\delta \rho(\tbf{x},t)]^{1/2} e^{i\varphi(\tbf{x},t)}$, 
where $\rho_0(\tbf{x}) = |\phi(\tbf{x})|^2$.  
The effective action for $\delta \rho$ and $\varphi$ to 
quadratic order is
\bea
S_{\mathrm{eff}} [\delta \rho, \varphi] &=&
\int dt \int d^2 \textbf{x} \mathscr{L}(\textbf{x},t)\, , \nonumber \\
\mathscr{L}&=&- \hbar \delta \rho \partial_t \varphi
                            -\frac{1}{8}\hbar^2 (\vec{\nabla}
			    \frac{\delta \rho}{\sqrt{\rho_0} })^2 
                            -\frac{1}{2} \rho_0 \hbar^2  (\vec{\nabla}
			    \varphi)^2 \nonumber \\ 
                            &-&\frac{1}{2} \int d^2 \textbf{x}'
			    V(\textbf{x}-\textbf{x}') \delta \rho(\textbf{x},t) \delta \rho_
			    (\textbf{x}', t) \, , \,~~
\eea
where $\rho_0(\textbf{x}) $ is the modulus square of the condensate wavefunction.

\begin{figure}
\includegraphics[width=0.5\linewidth]{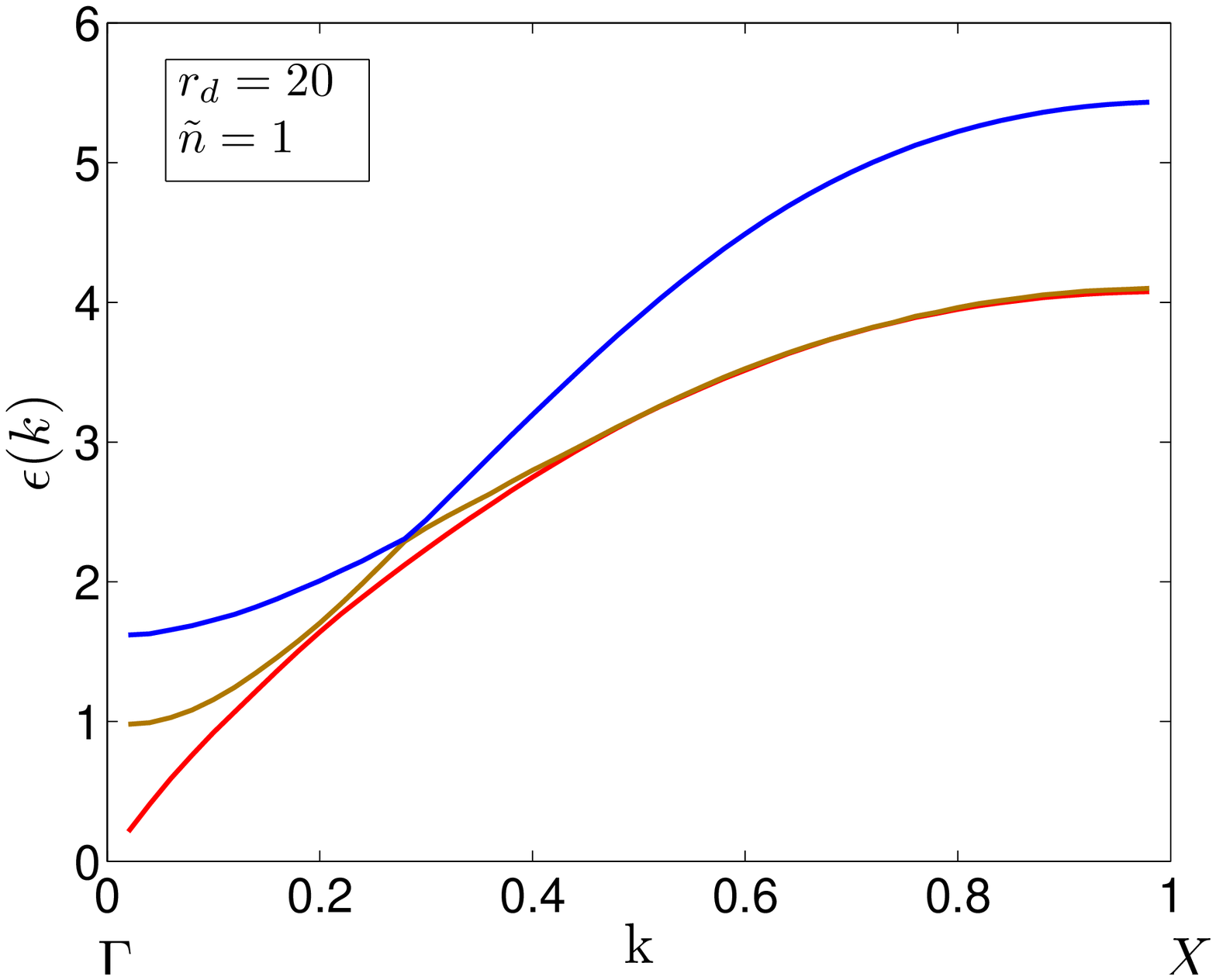}\includegraphics[width=0.5\linewidth]{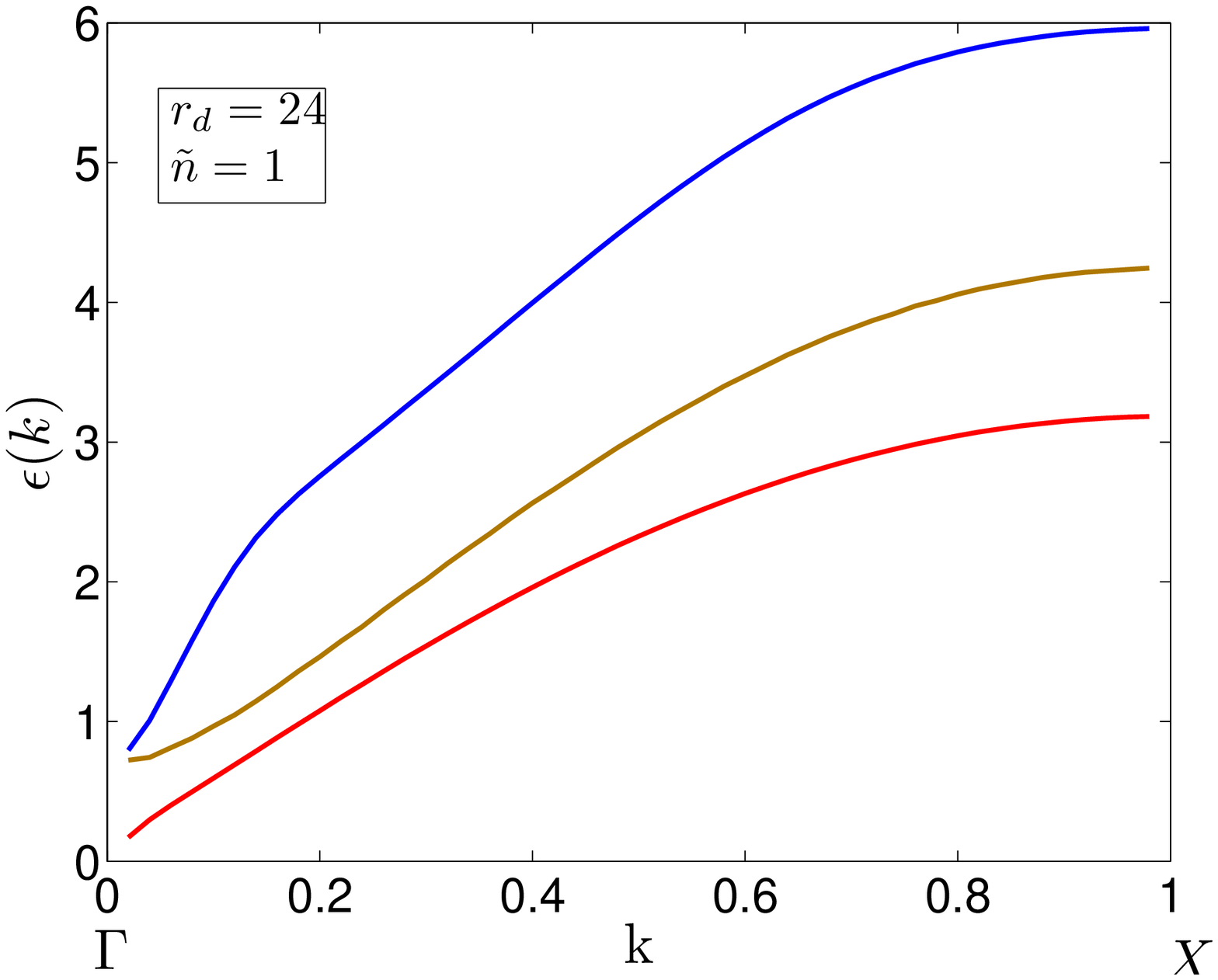}
\caption{{ The excitation spectrum on top of the BES state from 
the $\Gamma(\equiv(0,0))$ point to $X(\equiv(\frac{\pi}{a_{\text{BES}}},0))$ with $a_{\text{BES}}$ the lattice constant of the supersolid lattice.
The energy unit  is $\hbar^2/(m r_0^2)$.
There are branches of spectrum, the lowest three of which are shown above.} }
\label{BESSpectrum}
\end{figure}

%To study the low energy excitation spectrum, it is convenient to use the fields
% $\varphi(\textbf{x})$
%and $\eta(\textbf{x})\equiv \frac{\delta \rho(\textbf{x})}{\sqrt{ \rho_0}} $ 
% (instead of $\delta \rho(\textbf{x})$).
Since the effective theory possesses only discrete translational symmetries,
a Brillouin zone and its reciprocal lattice vectors can be defined. 
Let $\eta(\textbf{x})\equiv \frac{\delta \rho(\textbf{x})}{\sqrt{ \rho_0}} $. The effective action
in the momentum space is

\bea
&S_{\mathrm{eff}}&[\eta, \varphi]
 =\int dt \sum_{\textbf{k} \in {R}} \sum_{\textbf{K}_1, \textbf{K}_2 } 
	{\left[{A}_{\tbf{K}_1 \tbf{K}_2}(\tbf{k})
    \eta_{\tbf{K}_1+\tbf{k}} (\partial_t \varphi^{*}_{\tbf{K}_2+\tbf{k}}) \right. } \nonumber \\
                                   & +&\left. 
{B}_{\tbf{K}_1 \tbf{K}_2}(\tbf{k}) \eta_{\tbf{K}_1+\tbf{k}}
  \eta_{\tbf{K}_2+\tbf{k}} ^{*} + {C}_{\tbf{K}_1,\tbf{K}_2}(\tbf{k}) \varphi_{\tbf{K}_1+\tbf{k}}
  \varphi^{*}_{\tbf{K}_2+\tbf{k}}\right] \nonumber \\
	&+&c.c.\,  , \,~~
\label{s_eff}
\eea
with 
\bea
{A}_{\tbf{K}_1 \tbf{K}_2}(\tbf{k})& =& % \textstyle  
- { \alpha_{-\tbf{K}_1 +\tbf{K}_2}}, \nonumber \\
{B}_{\tbf{K}_1 \tbf{K}_2}(\tbf{k})& =& \textstyle  -\frac{1}{8}\hbar^2 (\tbf{K}_1 +\tbf{k})^2 \delta_{\tbf{K}_1, \tbf{K}_2} \nonumber \\ 
        && %\textstyle 
- \frac{1}{2} \sum_{\tbf{K}} \alpha_{\tbf{K}} \alpha_{\tbf{K}_2 -\tbf{K}_1 -\tbf{K}} U({\tbf{k}+\tbf{K}_2-\tbf{K}})  \, ,\nonumber \\
{C}_{\tbf{K}_1 \tbf{K}_2}(\tbf{k}) & =& % \textstyle  
\frac{1}{2} \sum_{\tbf{K}} {\hbar^2 [(\tbf{K}_1+\tbf{k})
  \cdot (-\tbf{K}_2 -\tbf{k})] \alpha_{\tbf{K}} 
     \alpha_{\tbf{K}_2 -\tbf{K}_1 -\tbf{K}} }\, , \, ~~
\eea 
%Here, $\tilde{A}_{\tbf{K}_1 \tbf{K}_2}(\tbf{k})= A_{\tbf{K}_1+\tbf{k}, -\tbf{K}_2-\tbf{k}}$, 
%$\tilde{B}_{\tbf{K}_1 \tbf{K}_2}(\tbf{k})=B_{\tbf{K}_1+\tbf{k}, -\tbf{K}_2-\tbf{k}}$,
%$\tilde{C}_{\tbf{K}_1 \tbf{K}_2}(\tbf{k})= C_{\tbf{K}_1+\tbf{k}, -\tbf{K}_2-\tbf{k}}$ with $\tbf{K}_1,\tbf{K}_2$ reciprocal lattice vectors. 
where $\tbf{K}_1, \tbf{K}_2 $ are reciprocal lattice vectors
and $\alpha_{\tbf{K}} $ is the Fourier component of $\sqrt{\rho_0(\tbf{x})}$.
The first Brillouin zone is divided into `{R}' (right) and `{L}' (left) subzones
according to time reversal; the summation $\tbf{k} \in {R}$ in 
Eq.~(\ref{s_eff})
 means summing 
over the `R' subzone.
The above effective theory is quadratic in
fields. Formally, the action can be written in a block diagonal form as:
$
S_{\mathrm{eff}}= \sum_{\tbf{k}\in {R}}
\left[ \begin{array} {cc}
\eta^{\dag} (\tbf{k}) & \varphi^{\dag}(\tbf{k})
\end{array} \right] \mathscr{G}_\tbf{k} ^{-1} 
\left[ \begin{array} {c}
\eta(\tbf{k}) \\ \varphi(\tbf{k})
\end{array} \right]
$
with 
\be
\mathscr{G}_\tbf{k} ^{-1} \equiv
\left[ \begin{array} {cc}
\tilde{B}(\tbf{k})^{T} +\tilde{B}(\tbf{k}) ^{ *} & \tilde{A}(\tbf{k})^{*} \partial_t \\
-\tilde{A}(\tbf{k})^{T} \partial_t & \tilde{C}(\tbf{k})^{T}+\tilde{C}(\tbf{k})^{*}
\end{array} \right] ,
\ee
where the crystal momentum $\tbf{k}$ is a good quantum number, 
and $\eta(\tbf{k})$ and $\varphi(\tbf{k})$ correspond to column vectors formed by 
$\{\eta_{\tbf{K}+\tbf{k}} \}$ and $\{\varphi_{\tbf{K}+\tbf{k}} \}$ with $\tbf{K}$ running over $p\tbf{G}_1 +q\tbf{G}_2$.
The energy spectrum is determined by the poles of 
$\mathscr{G}_\textbf{k}$, i.e., 
$\det [ \mathscr{G}_\textbf{k}^{-1}] =0$~\cite{Popov}. 
FIG.~\ref{BESSpectrum} shows typical results we obtained, 
which indicate the stability of the BES phase. 
However, for
sufficiently large $r_d$, the spectrum becomes imaginary near $\tbf{k}=\tbf{0}$
(crystal momentum) point
indicating instability of the BES state.  
%the BES state also becomes unstable. 
The stable regime of the BES state are shown in the phase diagram FIG.~\ref{PhaseD}.

\paragraph{Variationally compared with insulating crystal.}
We further estimate the energy of the IC state. The 
IC state is described by $ |\Psi_{\text{IC}}\rangle=\prod_{\vec{R}_i}
{c_{\vec{R}_i}^\dag}|0\rangle $ where the single particle wavefunction corresponding to the creation operator $c_{\vec{R}_i}^\dag$ is the Wannier
function $\phi_{\vec{R}_i}(\vec{r})$. Here we consider the case where each localized wave function
contains exactly one boson, forming a triangular lattice. The lattice constant $a_c$ is
thus determined by the density ( $a_c =[{2}/ (\sqrt{3} n)]^{1/2}$), which is different from 
the lattice constant of BES, $a_{\text{BES}}$, determined by the minimum point of $U(\tbf{k})$.
%, please give the right expression )}.  
The Wannier function is approximated by a localized Gaussian 
$\phi_{\vec{R}_i}(\vec{r}) \sim \frac{1}{\sqrt{\pi} \sigma} \exp(-\frac{(\vec{r}-\vec{R}_i)^2}{2\sigma^2})$
with $\sigma / a_c$ ranging from 0 to 0.3 over which the overlap between 
neighboring Gaussian wavefunctions can be neglected~\cite{Hartreevalid} and
the energy is obtained by $E_{\text{IC}} (\sigma)=\langle \Psi_{\text{IC}}| H | \Psi_{\text{IC}} \rangle$.
 When $\tilde{n} \gtrsim 1$, minimizing $E_{\text{IC}}(\sigma)$ gives $\sigma  /a_c \to 0.3$, indicating 
the IC state with the given lattice constant is unstable.  

%\bea
%&\langle \Psi_{IC}| H | \Psi_{IC} \rangle &= 
%&E_{IC}&=
%N \frac{\hbar^2}{2 m \sigma^2} \nonumber \\
%&+&\frac{N}{2} \sum_{\vec{R}_j \neq 0}\int d^2 x_1 d^2 x_2 V(x_1-x_2)
%\left \{|\phi_{0}(x_1)|^2 |\phi_{\vec{R}_j}(x_2)|^2 \right. 
%\nonumber \\
%&+&\left. \phi_0 (x_1) ^{*} \phi_{\vec{R}_j} (x_2)^{*} \phi_{\vec{R}_j} (x_1)  
% \phi_0 (x_2) \right \} \,  . \,~~
%&+&\left. \frac{N}{2} \sum_{\vec{R}_j \neq 0} \int d^2 x_1 d^2 x_2 \phi_0 (x_1) ^{*} \phi_{\vec{R}_j} (x_1) V(x_1 -x_2) 
%\phi_{\vec{R}_j} (x_2)^{*} \phi_0 (x_2) \right \}   
%do \, you\, have\, an\, expression?
%\eea

%We then compare the energy of this IC state with BES state and obtain 
%(lower) 
%phase boundary of BES state 
%(FIG.~\ref{PhaseD})  

In the phase diagram (FIG.~\ref{PhaseD}),  the BES is stable and is the most energy-favored 
in the `yellow shaded' regime. The lower boundary is determined by comparing the energy of
BES state and $E_{\text{IC}} (\sigma=0.3 a_c)$ while the right boundary 
is computed from the instability of the BES 
spectrum. In the `unstable' regime, the proposed BES state has lower energy than the IC state 
but is not stable against quantum fluctuations.

In conclusion, we studied a bosonic system with two-body interaction potentials
which display a negative minimum at a finite momentum and found a stable 
supersolid phase arising from Bose-Einstein condensation at finite momenta.  
The stability of
this novel supersolid phase is checked against quantum fluctuation.  A
unique feature of the BES state is that it breaks both
$U(1)$ and translational symmetry with a single order parameter,
namely, the superfluid order parameter $\langle
\psi\left(\vec{r}\right)\rangle$ is not only finite but also spatially
modulated.  The physical interpretation is that particles are not
localized in space but condensed to a single, common wavefunction
which is modulated like a solid.  This is conceptually different from
one of widely considered supersolid pictures of He-4~\cite{EKss,PO} 
 in which supersolidity is a mixture of two
orderings: atoms form charge-density-wave order (a crystal structure)
and in the same time vacancies or interstitials undergo usual
(zero-momentum) Bose-Einstein condensation.  For the conventional
superfluid phase originated from zero-momentum BEC, there exists long
range phase coherence but the phase correlation function is
homogeneous, not modulated, in space.  In contrast, for the IC (insulating
crystal) phase, particles are localized in space to each lattice site,
so there is no long range phase coherence. Therefore, as prediction
for cold gas experiments, a signature of the new BES phase
is the modulated phase coherence. This new state also opens fundamental
questions for future studies, for example, how the supercurrent is
affected by the simultaneous presence of crystalline ordering, and
topological configurations such as a vortex coupled to a crystal
defect. 

We thank Erhai Zhao and Uwe R. Fischer for helpful discussions. This work is supported
by ARO Grant No.  W911NF-07-1-0293.

\paragraph{Note Added.} Near the completion of this paper, a related,
independent study appeared~\cite{Zollerdroplet}, which discovered by exact
numerical algorithms for a similar model system a supersolid phase to
occur in the same parameter regime.

\bibliography{BECCrystal}

\begin{thebibliography}{20}
\bibitem[{not()}]{note1}
\bibinfo{note}{Another collective mode of the BES phase is lattice vibrations.
  All low energy fluctuations shall be studied together systematically in the
  future.}
\end{thebibliography}
\bibliographystyle{apsrev}

\end{document}